\documentclass[aps,amsmath,preprint,epsf,superscriptaddress,nofootinbib]{revtex4-1}
\usepackage{graphicx}
\usepackage{bm}
\usepackage{color}
\usepackage{amsmath,amssymb}	
\usepackage{hyperref}
\usepackage{setspace}
\usepackage{longtable}
\usepackage{booktabs}
\usepackage{comment}
\usepackage{array}
\usepackage{cancel}
\usepackage{enumitem}
\usepackage{physics}
\usepackage{mathtools}

\usepackage[english]{babel}
\usepackage[utf8x]{inputenc}
\usepackage[colorinlistoftodos]{todonotes}
\usepackage{slashed}

\def\slashchar#1{\setbox0=\hbox{$#1$} % set a box for #1
\dimen0=\wd0 % and get its size
\setbox1=\hbox{/} \dimen1=\wd1 % get size of /
\ifdim\dimen0>\dimen1 % #1 is bigger
\rlap{\hbox to \dimen0{\hfil/\hfil}} % so center / in box
#1 % and print #1
\else % / is bigger
\rlap{\hbox to \dimen1{\hfil$#1$\hfil}} % so center #1
/ % and print /
\fi}

\newcommand{\mpi}{\mathrm{\pi}}
\newcommand{\mpl}{M_{\mathrm{Pl}}}
\newcommand{\hc}{\mathrm{h.c.}}
\newcommand{\unit}[1]{\, \mathrm{#1}}

\begin{document}
\newcolumntype{Y}{>{\centering\arraybackslash}p{25pt}} 

%%%%%%%%%%%%%%%%%%%%%%%%%%%%%%%%%%
%%%%%%%%%%% Title page %%%%%%%%%%%
%%%%%%%%%%%%%%%%%%%%%%%%%%%%%%%%%%

\preprint{IPMU19-0121}

\title{A Dynamical Solution to the Axion Domain Wall Problem}

\affiliation{ICRR, The University of Tokyo, Kashiwa, Chiba 277-8582, Japan}
\affiliation{T. D. Lee Institute and School of Physics and Astronomy, Shanghai Jiao Tong University, Shanghai 200240, China}
\affiliation{Kavli IPMU (WPI), UTIAS, The University of Tokyo, Kashiwa, Chiba 277-8583, Japan}

\author{Masahiro Ibe}
\email[e-mail: ]{ibe@icrr.u-tokyo.ac.jp}
\affiliation{ICRR, The University of Tokyo, Kashiwa, Chiba 277-8582, Japan}
\affiliation{Kavli IPMU (WPI), UTIAS, The University of Tokyo, Kashiwa, Chiba 277-8583, Japan}
\author{Shin Kobayashi}
\email[e-mail: ]{shinkoba@icrr.u-tokyo.ac.jp}
\affiliation{ICRR, The University of Tokyo, Kashiwa, Chiba 277-8582, Japan}
\author{Motoo Suzuki}
\email[e-mail: ]{m0t@icrr.u-tokyo.ac.jp}
\affiliation{T. D. Lee Institute and School of Physics and Astronomy, Shanghai Jiao Tong University, Shanghai 200240, China}
\affiliation{ICRR, The University of Tokyo, Kashiwa, Chiba 277-8582, Japan}
\author{Tsutomu T. Yanagida}
\email[e-mail: ]{tsutomu.tyanagida@ipmu.jp}
\affiliation{T. D. Lee Institute and School of Physics and Astronomy, Shanghai Jiao Tong University, Shanghai 200240, China}
\affiliation{Kavli IPMU (WPI), UTIAS, The University of Tokyo, Kashiwa, Chiba 277-8583, Japan}

\date{\today}

%==========%
%	Abstract    %
%==========%
%=========================
\begin{abstract}
The domain wall problem and the isocurvature 
problem restrict possible combinations
of axion models and inflation models.
In this paper, we considered a new mechanism 
which solves those problems by 
dynamics of multiple scalar fields 
during/after inflation.
The mechanism makes axion models with a non-trivial domain wall number
compatible with inflation models 
with a large Hubble parameter, 
$H_{I} \gg 10^{7\mbox{--}8}$\,GeV.
The mechanism also avoids the isocurvature problem.
This mechanism 
increases the freedom of choice
of combinations of axion models
and inflation models.
\end{abstract}
%=========================

\maketitle

%=============%
%	Introduction    %
%=============%
%=========================
\section{Introduction}
\label{sec:introduction}
The Peccei-Quinn (PQ) mechanism is the most plausible solution to the Strong CP problem~\cite{Peccei:1977hh,Peccei:1977ur}.
In this mechanism, the effective $\theta$-angle of QCD is canceled by the vacuum expectation value 
(VEV) of the pseudo-Nambu-Goldstone boson, axion $a$, which is associated with the spontaneous
breaking of the global $U(1)$ symmetry (PQ symmetry)~\cite{Weinberg:1977ma,Wilczek:1977pj}.
The mechanism is particularly attractive as the invisible axion~\cite{Kim:1979if,Shifman:1979if,Zhitnitsky:1980tq,Dine:1981rt} 
is a good candidate for cold dark matter~\cite{Preskill:1991kd,Abbott:1982af,Preskill:1982cy}.

The domain wall problem and the isocurvature problem, however, restrict possible combinations
of axion models and inflation models.
For example, when spontaneous symmetry breaking of the PQ symmetry
takes place after the end of inflation, it triggers the formation of 
the cosmic (global) strings~\cite{Sikivie:1982qv}.
Around the cosmic string, the axion goes round its domain in $a/f_a = [0,2\pi N_{\rm DM})$.
Here, $f_a$ is the axion decay constant, and the integer $N_{\rm DM}\ge 1$ is the so-called domain wall number 
(see, e.g.~\cite{Kawasaki:2013ae,Kawasaki:2013iha}).
As the universe cools down below the QCD scale, the axion obtains a periodic scalar 
potential due to non-perturbative QCD effects, which leads to the formation 
of the axion domain wall around the cosmic string.
The formed string-wall network is stable unless $N_{\rm DM}=1$, which dominates over the energy density 
of the universe soon after its formation.%
\footnote{If the PQ symmetry is explicitly broken, the string-wall network is not exactly stable even 
for $N_{\rm DW} > 1$. 
However, the explicit breaking which is required to make the string-wall network collapses fast enough 
encounters the strong CP problem again~\cite{Hiramatsu:2012sc}. 
}

Until today, there are a few solutions to the domain wall problem.
For example, the trivial domain wall number, $N_{\rm DW} = 1$, is possible in the KSVZ model~\cite{Kim:1979if,Shifman:1979if}.
In this case, only one domain wall attaches to each of the cosmic string, and hence, 
the string-wall network collapses immediately after the QCD phase transition~\cite{Vilenkin:1982ks}.
As a notable feature of this scenario, the abundance of the axion dark matter is dominated 
by the contribution from the decay of the string-wall network,
and the observed dark matter density is explained for $f_a \sim 10^{10}$\,GeV~\cite{Hiramatsu:2012gg}.
The required decay constant is much smaller than $f_a \sim 10^{12}\,$GeV
which is appropriate for the so-called misalignment mechanism for the axion dark matter.
These scenarios can be distinguished by the axion search experiments (see, e.g. \cite{Graham:2015ouw}).%

Another possibility to evade the domain wall problem is 
to assume the PQ symmetry breaking before inflation.
In this case, the axion takes a single field value in our universe 
with a tiny quantum fluctuation, and hence,
no domain wall is formed below the QCD scale.
The quantum fluctuation of the axion, on the other hand, induces 
the isocurvature perturbations of cold dark matter. 
To avoid too large isocurvature perturbations, 
the Hubble parameter during inflation, $H_{I}$, must be smaller than about 
$10^{7\mbox{--}8}$\,GeV~\cite{Akrami:2018odb} (see also \cite{Kawasaki:2013ae,Kawasaki:2018qwp}).
This constraint significantly restricts the variety of inflation models.

In this paper, we discuss a new mechanism which solves those problems by dynamics of multiple scalar fields during/after inflation.
The mechanism makes the axion models with a non-trivial domain wall number
compatible with the inflation models with a large Hubble parameter, 
$H_{I} \gg 10^{7\mbox{--}8}$\,GeV.
The mechanism also avoids the isocurvature problem.
This mechanism increases the freedom of choice
of combination of axion models
and inflation models. 
In particular, this mechanism makes 
the axion model with $f_a \sim 10^{12}$\,GeV
compatible with the models of inflation with $H_I\sim 10^{13}$\,GeV.
Such a large axion decay constant and the large Hubble parameter during inflation 
can be tested by the axion search experiments (see, e.g. \cite{Graham:2015ouw}) and the searches for the primordial 
$B$-mode polarization in the cosmic microwave background (CMB) (see, e.g. \cite{Abazajian:2016yjj}), respectively.

In the new mechanism, we may consider 
any type of the axion model.
We call the PQ charged field which spontaneously breaks the PQ symmetry, the PQ field.
Then, we introduce an additional PQ charged scalar field which obtains a vanishing VEV.
We call this additional field the spectator PQ field.
We assume that the spectator PQ field obtains a large field value during/after inflation.
The large field value of the spectator PQ field provides a non-trivial scalar potential 
of the axion when the PQ field obtains its VEV after inflation.
The non-trivial axion potential prohibits the formation of the cosmic string, and hence,
prohibits the string-wall network below the QCD scale.
The isocurvature perturbation which stems from the quantum fluctuation of 
the spectator PQ field is suppressed by its large field value during inflation~\cite{Linde:1991km}.
In this way, the new mechanism solves the domain wall problem without causing the isocurvature 
problem.
This mechanism may be regarded as 
a multi-field version of the mechanism discussed in \cite{Harigaya:2015hha}.%
\footnote{See also \cite{Takahashi:2015waa,Kawasaki:2015lpf,Kearney:2016vqw,Ho:2018qur} for other realization of dynamics 
which solves the domain wall and the isocurvature problems.
The solutions in the context of the axion predicted in the string theory~\cite{Witten:1984dg,Kallosh:1995hi,Svrcek:2006yi} 
have also been discussed, where the axion dark matter abundance is suppressed dynamically~\cite{Kawasaki:2017xwt,Co:2018phi}.
}

The organization of the paper is as follows.
In section~\ref{sec:model setup}, we summarize the setup of our model.
In section~\ref{sec:dynamics of S}, we discuss how
the spectator PQ field evolves.
In section~\ref{sec:Dynamics of Axion}, we discuss
how the axion behaves in the presence of the spectator 
PQ field.
We also discuss viable parameter region of the new mechanism.
In section~\ref{sec:SUSY}, we discuss supersymmetric 
extension of the model.
The final section is devoted to our conclusions.

%=============%
%	Model Setup    %
%=============%
%=========================
\section{Peccei-Quinn mechanism with a spectator PQ field}
\label{sec:model setup}
\subsection{General Recipe of the Dynamical Solution}
Before discussing the details of the mechanism, 
we summarize the general recipe 
for the dynamical solution of the domain wall and the isocurvature problems.
\begin{enumerate}[itemsep=1mm]
    \item Bring an axion model where the PQ symmetry is spontaneously broken by the VEV of the PQ field, $P$.
    \item Add a spectator PQ field, $S$, 
    which obtains a vanishing VEV
    but has a large field value in the early universe until 
    $P$ obtains the VEV (section \ref{sec:dynamics of S}).
    \item Introduce a mixing term between $P$ and $S$ 
    so that $P$ feels a strong PQ symmetry breaking effects 
    when it obtains the VEV (section \ref{sec:Dynamics of Axion}).
    \item Make the effects of the mixing term inefficient before $S$  starts coherent oscillation around its origin
    (section \ref{sec:Dynamics of Axion}).
\end{enumerate}
With the large field value of $S$, 
$P$ feels a strong PQ symmetry breaking,
and no cosmic strings are formed when $P$ obtains the VEV. 
The fourth condition is important not to randomize the axion field value even after $S$ starts the coherent oscillation
(subsection~\ref{sec: Axion ramdomize}).
Without cosmic strings and with the uniform axion field value, the domain walls are not formed after the QCD phase transition.
The quantum fluctuation of the phase component of $S$ is imprinted in the axion through the mixing term.
The isocurvature problem can be avoided by requiring that 
the field value of $S$ is of $\order{\mpl}$ during inflation
(subsection~\ref{sec:Isocurvature Perturbation of Axion}).
In any successful implementation of this mechanism, 
the domain wall and the isocurvature problems are solved dynamically.

\subsection{KSVZ Axion Model}
As a concrete example of the axion model,
we consider the KSVZ axion model~\cite{Kim:1979if,Shifman:1979if}
in which the PQ field, $P$, obtains the VEV via the scalar potential,
\begin{align}
   V(P) = \lambda_p\qty(\abs{P}^2-\frac{v_{\mathrm{PQ}}^2}{2})^2\ .
\end{align}
Here, $\lambda_p$ is a coupling constant of ${\cal O}(1)$ and $v_{\mathrm{PQ}}$
is a parameter with mass dimension.
The VEV of the PQ field is given by, $\langle P\rangle = v_{\rm PQ}/\sqrt{2}$.
The axion field, $a$, corresponds to the phase component of $P$, \begin{align}
\label{eq:axion}
    P = \frac{1}{\sqrt 2} v_{\mathrm{PQ}} \, e^{{i a}/v_{\mathrm{PQ}}}\ ,
\end{align}
where we omit the radial component of $P$ for brevity.

The PQ field couples to the $N_f$ vector-like quarks in
the fundamental representation of the $SU(3)$ 
gauge group of QCD, $(Q_L,  \bar{Q}_R)$ via
\begin{align}
\label{eq:KSVZquark}
    {\cal L} =  y_{\rm KSVZ} P Q_L{\bar Q_R} + h.c.\ ,
\end{align}
with $y_{\rm KSVZ}$ being the coupling constant.
Below the mass scale of the KSVZ quarks, $y_{\rm KSVZ}v_{\rm PQ}$, 
the QCD anomaly induces the axion couplings to QCD,
\begin{align}
\label{eq:QCD}
  L= \frac{g_s^2}{32\pi^2 } \frac{N_f}{v_{\mathrm PQ}}\,a\,
  G\tilde{G}\ ,
\end{align}
Here, $g_s$ denotes the QCD gauge coupling constant
and $G$ and $\tilde G$ are the QCD field strength 
and its hodge dual, respectively.
The Lorentz and color indices are understood. 
We define the origin of the axion field space at which the effective $\theta$-angle of QCD is vanishing.

Below the QCD scale, the above interaction term in 
Eq.\,\eqref{eq:QCD} leads to the scalar potential
of the axion,
\begin{align}
    V(a) \sim m_a^2 f_a^2 \left[1-\cos\frac{a}{f_a}\right]\ .
\end{align}
Here, $f_a = v_{\mathrm PQ}/N_f$ is the effective decay constant of the axion, and $m_a$ denotes the mass of the axion which is estimated to be,
\begin{align}
    m_a \simeq 6\,\mu\mathrm{eV}\left(\frac{10^{12}\,\rm GeV}{f_a}\right)\ ,
\end{align}
(see e.g. \cite{Wantz:2009mi,diCortona:2015ldu}).
It should be noted that the domain of the axion field is given by
$a/f_a = [0,2\pi N_{f})$, and hence, $N_f$ corresponds to the domain wall number, $N_{\rm DW} = N_f$.

As we will see below, the axion field value settles 
to a non-zero value $a_i$ of ${\cal O}(f_a)$ 
after a complex dynamics of the new mechanism.
Below the QCD scale, the axion starts coherent oscillation from 
the non-zero field value around its origin which behaves 
as cold dark matter as in the conventional 
misalignment mechanism~\cite{Preskill:1991kd,Abbott:1982af,Preskill:1982cy}.
The axion dark matter density is given by~\cite{Bae:2008ue},
\begin{align}
\Omega_a h^2  \simeq 0.2 \times \left(
\frac{a_i}{f_a}\right)^2
\left(
\frac{f_a}{10^{12}\,\rm GeV}
\right)^{1.19}\ .
\end{align}
Based on this estimate, we focus on the case with 
$f_a = {\cal O}(10^{12})$\,GeV in the following discussion.

\subsection{Spectator PQ Field}
Now let us introduce another PQ charged scalar field,
the spectator PQ field, $S$.
We assume that $S$ has a PQ charge 
which is $-1/m$ of that of the PQ 
field $(m \in {\mathbb N})$. 
With this assumption, $S$ couples to $P$ via,
	\begin{align}
		\label{eq: Potential of P,S}
		V(P,S)=&\lambda_p\qty(\abs{P}^2-\frac{v_{\mathrm{PQ}}^2}{2})^2+m_S^2\abs{S}^2+\frac{1}{(n!)^2}\frac{\lambda_s^2}{\mpl^{2n-4}}\abs{S}^{2n}+
		\frac{\lambda}{m!\mpl^{m-3}}S^mP+\hc
	\end{align}
Here, $m_S$ is the mass parameter of $S$, $\lambda_s$ and $\lambda$ are dimensionless coupling constants, 
and $\mpl\simeq 2.4\times 10^{18}$\,GeV 
the reduced Planck scale.
Due to the positive mass squared, 
$S$ does not obtain a non-vanishing VEV.
As we will see in the following two sections, 
$n$ is required to be larger than $5$
for a successful mechanism.
The absence of the lower dimensional scalar potential terms
of $S$ than $|S|^{2n}$ will be justified in 
the supersymmetric extension discussed 
in section~\ref{sec:SUSY}.

In this mechanism, a large field value 
of $S$ during/after inflation plays a 
crucial role to solve the domain wall problem
and the isocurvature problem.
For that purpose, we introduce interactions 
between the (spectator) PQ fields with  the inflaton field $\phi$,
	\begin{align}
		\label{eq: Potential of P,S and inflaton}
		V(P,S,\phi)=&V(\phi)+V(P,S)+\frac{c_p}{3}\frac{V({\phi})}{\mpl^2}\abs{P}^2		%
		-\frac{c_s}{3}\frac{V({\phi})}{\mpl^2}\abs{S}^2 \ ,
	\end{align}
where $c_p$ and $c_s$ are positive valued coupling constant.
$V(\phi)$ denotes the inflaton potential with which the Hubble parameter during inflation is given by,
\begin{align}
    H_{I}^2= \frac{V(\phi)}{3M_{\rm Pl}^2}\ .
\end{align}
Through the interactions with the inflaton, $P$ and $S$ 
obtain the positive and the negative Hubble-induced mass terms during inflation,
\begin{align}
\label{eq:HubbleMass}
    \tilde{m}_{P}^{2} = c_p H_{I}^2 \,(>0) \ , 
    \quad & \tilde{m}_{S}^2 = - c_s H_I^2 \,(<0)\ ,
\end{align}
respectively.
Generally, scalar fields obtain Hubble-induced mass terms.
We also discuss how the interactions with the inflaton 
in Eq.\,\eqref{eq: Potential of P,S and inflaton} can be obtained 
in the supersymmetric extension.

In the following scenario, we assume that $S$ is never 
in the thermal equilibrium.
Such a situation can be easily realized 
when the inflaton field mainly decays into the Standard Model particles.
Late time interactions of $S$ with thermal bath particles 
are negligible as it only couples to other fields 
through the Planck suppressed operators.

%==============%
%	Dynamics of S   %
%==============%
%=========================
\section{Dynamics of the spectator PQ field}
\label{sec:dynamics of S}
	%============%
	%	Inflation Era  %
	%============%
	%=========================
\subsection{Inflation Era}
\label{subsec:Inflation Era}
During inflation, $S$ obtains a negative 
Hubble-induced mass term in Eq.\,\eqref{eq:HubbleMass},
which is much larger than $m_S^2$ in size.
Thus, the potential of $S$ can be approximated by
\begin{align}
	\label{eq:Potential of S in inflation era}
		V(S)=\frac{\lambda_s^2}{(n!)^2\mpl^{2n-4}} \abs{S} ^{2n}-c_sH_I^2\abs{S}^2\ .
\end{align}
Due to the Hubble-induced mass term,
$S$ obtains a large expectation value,
\begin{align}
		\label{eq:VEV of S in inflation era}
			\langle {S_I} \rangle \simeq\qty(\sqrt{\frac{c_s}{n}}\frac{n!}{\lambda_s})^{\frac{1}{n-1}}\qty(\frac{H_I}{\mpl})^{\frac{1}{n-1}}\mpl\ .
\end{align}
Hereafter, the expectation value, $\langle S\rangle$, denotes its absolute value if not otherwise specified.
In the following analysis, we assume  
\begin{align}
    \lambda_s \gtrsim n!\sqrt{\frac{c_s}{n}}
    \left(\frac{H_I}{\mpl}\right)\ ,
\end{align}
so that $\langle S_I\rangle$ not to exceed ${\cal O}(\mpl)$.
For example, 	$\langle {S_I} \rangle\sim \mpl$, 
for $n=6$, $H_{I}=10^{12}$\,GeV, $c_s = 1$ 
	and $\lambda_s=10^{-4}$. 
	
	%==============%
	%	Oscillation Era  %
	%==============%
	%=========================
\subsection{Inflaton Oscillation Era}
\label{subsec:Oscillation Era}
After inflation, the inflaton starts coherent oscillation
around its minimum.
As the inflaton oscillation time scale
becomes much shorter than the Hubble time,
the dynamics of $S$ 
can be analyzed by taking the time average
of the inflaton oscillation.
Thus, the inflaton potential in Eq.\,\eqref{eq: Potential of P,S and inflaton} can be approximated by 
\begin{align}
	\overline{V(\phi)}=\frac{3}{2}H^2\mpl^2\ ,
\end{align}
where the bar denotes the time average and 
$H$ is the Hubble parameter at that time.\footnote{At the beginning of the inflaton oscillation, in the case of chaotic inflation, the time scale of the inflaton oscillation is comparable to the Hubble time. However, this does not change the dynamics of $S$ significantly.}

Now, let us focus on the dynamics of the radial component 
of $S$,
\begin{align}
	S=\frac{\chi}{\sqrt{2}}\ , \quad (\chi \in {\mathbb R}) \ .
\end{align}
Due to the PQ symmetry, the potential of $S$ does not induce the torque in the complex plane of $S$.
Thus, the motion of $S$ is confined on a straight line  passing through $S=0$ 
(see e.g. Fig.~\,\ref{fig:phasemotion}).
We will discuss how the phase component of $S$ behaves in the next section.
In the above approximation, the equation of motion (EOM) of the zero-mode of $\chi$ is given by,
\begin{align}
	\label{eq: EOM of chi}
	\ddot{\chi}+3H\dot{\chi}+\frac{n\lambda_s^2}{2^{n-1}(n!)^2\mpl^{2n-4}} \chi ^{2n-1}-\frac{c_s}{2}H^2\chi=0 \ ,
\end{align}
where the dot denotes the time derivative.
We neglect $m_s^2$ by assuming that it is still much smaller 
than $H^2$ in this period.

Following \cite{Harigaya:2015hha}, we introduce the $e$-folding number,
$N \equiv \ln R$, as a time variable, where $R$ is the scale factor of the universe.
We define $R$ such that $R =1$ when the inflaton starts to oscillate. 
With the $e$-folding number, the EOM is rewritten by,
\begin{align}
	\dv[2]{N}\chi+\frac{3}{2}\dv{N}\chi+\frac{n\lambda_s^2}{2^{n-1}(n!)^2\mpl^{2n-4}H^2}\chi^{2n-1}-\frac{c_s}{2}\chi=0\ .
\end{align}
	Next, we set $\chi$ in the form of 
		\begin{align}
		\label{eq:sigma}
			\chi=\sigma\sqrt{2}\mpl\qty(\frac{c_s(n!)^2H_i^2}{2n\lambda_s^2\mpl^2})^{\frac{1}{2(n-1)}}\exp\qty[-\frac{3N}{2(n-1)}]\ .
		\end{align}
Here, $H_i$ denotes the Hubble parameter at the on-set of the inflaton oscillation.
This leads to the EOM of $\sigma$
		\begin{align}
		\label{eq:EOM}
			\dv[2]{N}\sigma+\frac{3(n-3)}{2(n-1)}\dv{N}\sigma+\frac{c_s}{2}\sigma^{2n-1}-\qty(\frac{9(n-2)}{4(n-1)^2}+\frac{c_s}{2})\sigma=0\ .
		\end{align}
		
The EOM of $\sigma$ represents a motion of a particle in a potential,
\begin{align}
    V(\sigma) =- \frac{1}{2} \left(\frac{9(n-2)}{4(n-1)^2}+\frac{c_s}{2}\right) \sigma^{2} + \frac{c_s}{4n} \sigma^{2n}\ ,
\end{align}
which has a minimum at,
\begin{align}
	\label{eq:sigma0}
		\sigma_0=\qty(1+\frac{9(n-2)}{2(n-1)^2c_s})^{\frac{1}{2(n-1)}}\ .
\end{align}
The second term of the EOM is a velocity dependent force.
The initial position of $\sigma$ is given by Eq.\,\eqref{eq:VEV of S in inflation era},
\begin{align}
\label{eq:init}
    \sigma_i \sim 2^{\frac{1}{2(n-1)}}\ , 
\end{align}
which is close to $\sigma_0$ for $c_s = {\cal O}(1)$.

The velocity dependent force in Eq.\,\eqref{eq:EOM}
plays the role of friction for $n\ge 4$ (see also \cite{Harigaya:2015hha}).
Thus, for $n\ge 4$, $\sigma$ which starts oscillation from $\sigma_0$ immediately 
settles down to $\sigma_0$ and stays there during the inflaton oscillation era.
In this case,
the field value of $S$ is roughly given by
\begin{align}
    \label{eq:VEV of S in oscillation era}
	\langle{S}\rangle\simeq\qty(\sqrt{\frac{c_s}{n}}\frac{n!}{\lambda_s})^{\frac{1}{n-1}}\qty(\frac{H}{\mpl})^{\frac{1}{n-1}}\mpl\ ,
\end{align}
which has the same dependency on the Hubble parameter 
in Eq.\,\eqref{eq:VEV of S in inflation era}.
This behavior of the scalar field is called as the scaling solution~\cite{Liddle:1998xm}.

For $n=3$, $\sigma$ keeps oscillating around $\sigma_0$ 
but does not go over $\sigma = 0$~\cite{Harigaya:2015hha}.
Thus, in this case, $S$ again keeps a large field value during the inflaton oscillation era.%
\footnote{The parametric resonance due to the oscillation of $S$ is not effective~\cite{Harigaya:2015hha}.}
As pointed out in \cite{Ema:2015dza}, however, the scalar field shows a peculiar behaviour for $n=3$, the pseudo-scaling solution, in which the field value gradually decreases in a zigzag manner.

For $n<3$,  the velocity dependent force accelerates the motion of $\sigma$,
and eventually, the oscillation of $S$ goes over $S=0$,
which results in $\langle S \rangle = 0$.
For a successful solution to the domain wall 
problem, we require that $S$ has a large field value 
when $P$ obtains the VEV in the radiation dominated (RD) era. 
Thus, we at least require $n \ge 3$ so that $S$ keeps a large field value during the inflaton oscillating period.
As we will immediately see, however, we eventually require $n \ge 5$ for $S$ to have a large field value 
in the RD era (see also \cite{Liddle:1998xm}).

	%=========================
	
	%=============%
	%	Radiation Era  %
	%=============%
	%=========================
\subsection{Radiation Dominated Era}
	\label{subsec:Radiation Dominated Era}
In the RD era, the time dependence of the Hubble parameter changes from that in the inflaton oscillation era.
Besides, the PQ fields no longer obtain the Hubble-induced mass terms through the interactions with the inflaton.

In this mechanism, we assume that $m_s$ is much smaller than $H$ 
at the beginning of the RD era.
Then, the EOM of $S$ can be written as
\begin{align}
	\ddot{\chi}+3H\dot{\chi}+\frac{n\lambda_s^2}{2^{n-1}(n!)^2\mpl^{2n-4}} \chi ^{2n-1}=0\ .
\end{align}
As in the inflaton oscillation era, we introduce the $e$-folding number $N\equiv\ln R$ 
which is vanishing at the beginning of the RD era. 
By using the $e$-folding number, the EOM is rewritten by,
\begin{align}
	\dv[2]{N}\sigma+\frac{n-5}{n-1}\dv{N}\sigma+\sigma^{2n-1}-\frac{2(n-3)}{(n-1)^2}\sigma=0 \ .
\end{align}
Here, we set 
\begin{align}
	\chi=\sigma\sqrt{2}\mpl\qty(\frac{(n!)^2H_{r,i}^2}{n\lambda_s^2\mpl^2})^{\frac{1}{2(n-1)}}\exp\qty[-\frac{2N}{n-1}]\ ,
\end{align}
with $H_{r,i}$ being the Hubble parameter at the beginning of the RD era.
	
As in the case of the inflaton oscillation era, 
the EOM corresponds to a participle motion in a potential
\begin{align}
    V(\sigma) = - \frac{(n-3)}{(n-1)^2} \sigma^2 + \frac{1}{2n} \sigma^{2n}\ , 
\end{align}
which has the minimum at,
\begin{align}
	\sigma_{r,0}=\qty[\frac{2(n-3)}{(n-1)^2}]^{\frac{1}{2(n-1)}}\ .
\end{align}
The initial position of $\sigma$ is roughly given by,
	\begin{align}
	\label{eq:sigmari}
	    \sigma_{r,i} \sim c_s^{\frac{1}{2(n-1)}}\ .
	\end{align}

Similarly to the case of inflaton oscillation era, the friction term has a wrong sign for $n\leq 4$.
For $n>5$, on the other hand, $S$ again behaves as the scaling solution~\cite{Liddle:1998xm},
\begin{align}
\label{eq:Behavior of S}
	\langle{S}\rangle\simeq \qty[\frac{2(n-3)(n!)^2}{n(n-1)^2\lambda_s^2}]^{\frac{1}{2(n-1)}}\qty(\frac{H}{\mpl})^{\frac{1}{n-1}}\mpl
	\simeq\qty(\frac{n!}{\lambda_s})^{\frac{1}{n-1}}\qty(\frac{H}{\mpl})^{\frac{1}{n-1}}\mpl\ .
\end{align}
For $n=5$, $S$ behaves as the pseudo-solution~\cite{Ema:2015dza}.
In summary, $S$ keeps a large field value during the RD era for $n\ge 5$.
In the following analysis, we take $n=6$ as the minimal model, 
since the zigzag behavior in the pseudo scaling solution makes the
analysis complicated.

While $H\gtrsim m_S$, $S$ obeys the scaling solution.\footnote{When $S$ obeys the scaling solution, $S$ does not lose its potential energy through particle emission.}

When the Hubble parameter decreases further and becomes smaller than $m_s$,
$S$ finally starts to oscillate around its origin.
For $n=6, \lambda_s=10^{-4}$ and $m_S \lesssim 10^{-15} \unit{GeV}$, the contribution of the coherent oscillation of $S$ to the DM abundance is negligibly small.\footnote{For $m_S =\order{10^{-15}}\unit{GeV}$, the coherent oscillation of $S$ can be the dominant component of DM. Interestingly, this case provides an ultra-light DM model whose initial condition of the coherent oscillation is dynamically determined. }
For a larger $m_S$, we assume that $S$ decays into massless fermions 
$\psi_s$ through
	\begin{align}
	\label{eq:Decay channel of S}
		\mathcal{L}_{S\mbox{-}\mathrm{decay}}=-y_sS
		\bar{\psi_s}\psi_s+\hc
	\end{align}
With this assumption, the energy density of $S$
does not cause any cosmological problem.
The number density of the massless fermions is also 
negligibly small and does not contribute to the 
dark radiation (see the Appendix~\ref{sec:energy density of S}).

As we will see below, the behavior of $S$ discussed in this section 
successfully solves the domain wall problem and the isocurvature problem.

%=================%
%	Dynamics of Axion   %
%=================%
%=========================
\section{Dynamics of Axion and Constraints}
\label{sec:Dynamics of Axion}
Now, let us consider the dynamics of the PQ field and the axion.
A notable feature of this mechanism is that the large field value of $S$ provides a non-vanishing effective 
linear term of $P$ through the $P$-$S$ mixing term 
(see Eq.\,\eqref{eq: Potential of P,S and inflaton}).

%==================%
%	Before PQ breaking  %
%==================%
%=========================
\subsection{Before PQ Breaking}
\label{subsec:Before PQ Breaking}
During inflation, the minimum of the scalar potential of $P$ is shifted due to the effective linear term, and $P$ also obtains a non-zero field value which is determined 
by ballancing between the quartic term $\lambda_p\abs{P}^4$ 
and the effective linear term $\lambda \langle S \rangle^m P$, which is given by 
\begin{align}
\label{eq:P vev inflation}
	\langle P_I \rangle &\sim \mpl\qty(\frac{\lambda}{m!\lambda_p})^{1/3}\qty(\frac{\langle S_I\rangle}{\mpl})^{m/3} \notag \\
	&\simeq \mpl\qty[\frac{\lambda}{m!\lambda_p}\qty(\frac{n!}{\lambda_s})^{\frac{m}{n-1}}]^{1/3}\qty(\frac{H_I}{\mpl})^{\frac{m}{3(n-1)}}\ .
\end{align}
As we will see in the following, the minimal model 
which successfully solves the domain wall problem  
is the one with $n=6$, $m=11$.
In this case, $\langle P_I \rangle$ is rather large 
during inflation,
\begin{align}
\label{eq:P vev inflation num}
	\langle P_I \rangle \sim 10^{14}\unit{GeV}\ ,
\end{align}
where the parameters are set to $\lambda_p=1$, $\lambda_s=10^{-4}, \lambda=10^{-6}$ 
and $H_I=10^{12}\unit{GeV}$ as a benchmark point (see section~\ref{sec: constraints}).
Accordingly, the KSVZ quarks obtain heavy masses through
the Yukawa coupling in Eq.\,\ref{eq:KSVZquark},
\begin{align}
	M_{KSVZ}\simeq y_{\mathrm{KSVZ}}\langle P_I \rangle\ ,
\end{align}
during inflation.
	
In the inflaton oscillation era, on the other hand,
$S$ decreases as the scaling solution, 
and hence, the minimum of the potential of $P$ also decreases.
By the time of the completion of reheating process,
the minimum position of $P$ becomes much smaller than 
the temperature of the universe, $T$.%
\footnote{Here, $P$ does not necessarily follow the minimum 
of the potential, although such a behavior does not affect 
the following argument.}
Therefore, the KSVZ quarks and $P$ are thermalized by the completion of the reheating process.
Once the KSVZ quarks and $P$ are thermalized, $P$ obtains a thermal potential.\footnote{$S$ and $P$ mix through $S^mP$ term, however the mixing is not large enough for $S$ to be thermalized. }
	
Due to the thermal mass of $P$ of ${\cal O}(T^2)$,%
\footnote{We assume $\lambda_p$ and $y_{\rm KSVZ}$ 
are of ${\cal O}(1)$.}
$P$ is settled to
\begin{align}
	\langle P \rangle &\sim \frac{\lambda}{m! T^2\mpl^{m-3}}\langle S\rangle^m \notag \\
	 &\simeq \frac{\lambda}{m!}\qty(\frac{n!}{\lambda_s})^{\frac{m}{n-1}}\qty(\frac{T}{\mpl})^{\frac{2m}{n-1}-2}\mpl\ .
\end{align}
This expectation value is much smaller than $v_{\rm PQ} = {\cal O}(10^{12})$\,GeV.
Thus, we can consider that $P$ is stabilized at the origin 
by the thermal mass term at the beginning of the RD era.

%=========================
	
%================%
%	After PQ breaking  %
%================%
%=========================
\subsection{After PQ Breaking}
\label{subsec:After PQ Breaking}
\begin{figure}
	\centering
	\includegraphics[width=0.35\linewidth]{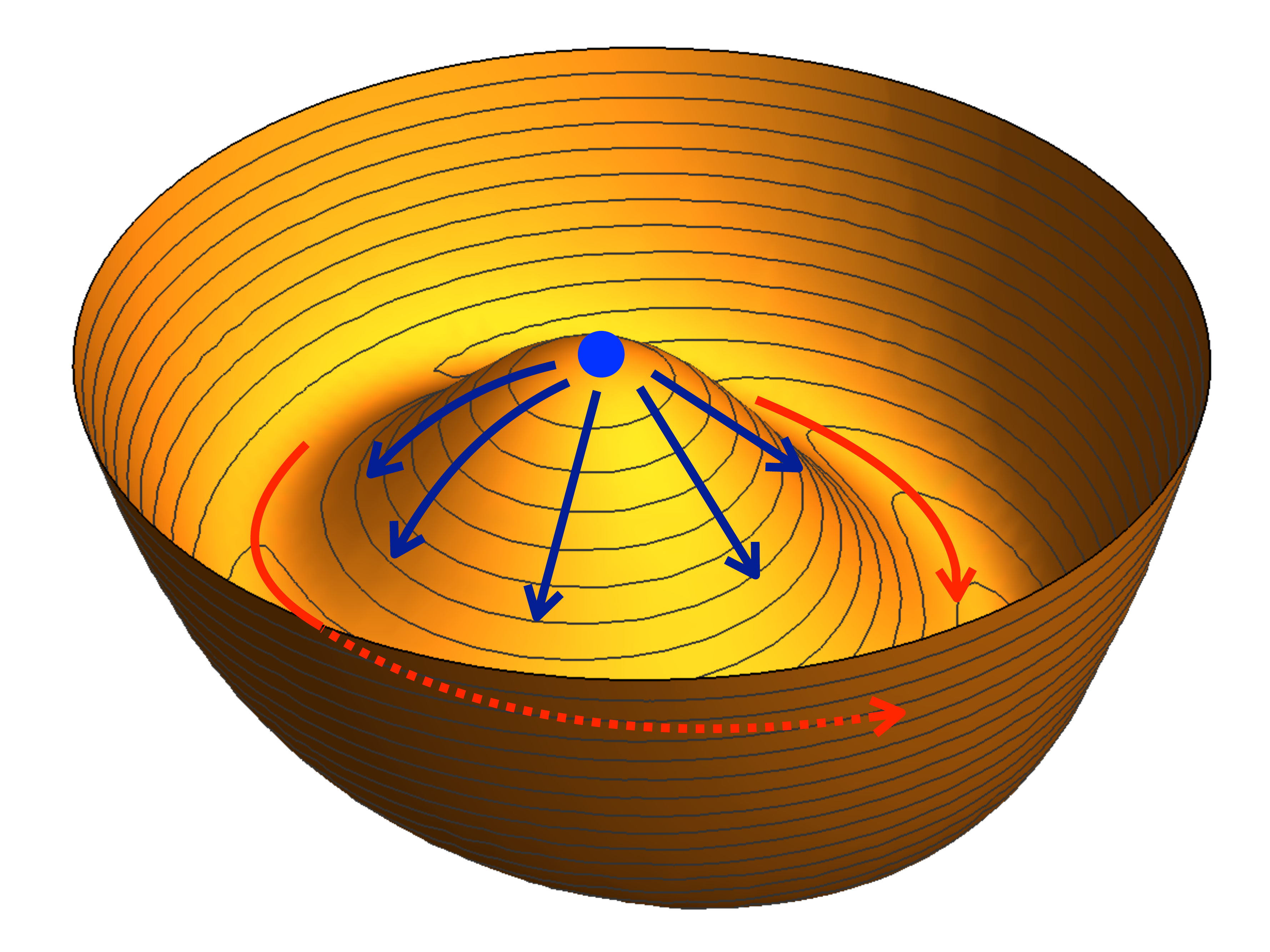}
	\caption{A schematic picture of the axion evolution when 
	$P$ obtains the VEV.
	The tilted Mexican hat corresponds the scalar potential of the PQ field
	on a complex plane of $P$.
	In each Hubble patch, the axion rolls down from 
	the hilltop of the potential in a random direction.
	The axion settles to a unique field value due 
	to the bias term made by the large field value of $S$.}
	\label{fig:axion}
\end{figure}

When $P$ obtains the VEV, $\langle P \rangle = v_{\rm PQ}/\sqrt{2}$, 
the axion rolls down from the hilltop of the potential.
The direction of the axion is random in
each Hubble patch~(see Fig.\,\ref{fig:axion}), which 
results in the formation of the cosmic strings.
However, the axion settles down to its minimum of the cosine potential induced by the $P$-$S$ mixing term,
which forms the domain walls around the strings.
This situation is analogous to the string-wall network formation below the QCD scale in the $N_{\mathrm{DW}}=1$ scenario.
In $N_{\mathrm{DW}}=1$ scenario, the string-wall network collapses by itself when 
the energy of a domain wall exceeds that of a string~\cite{Vilenkin:1982ks}.

The condition for the collapse of the string-wall network is given by
\begin{align}
	\label{eq:Domain wall collapsing condition}
	\frac{\sigma_{\mathrm{w}}d_H^2}{T_{\mathrm{s}}d_H}>1\ ,
\end{align}
where $\sigma_{\mathrm{w}}$ is the surface tension of the domain wall, 
$T_{\mathrm{s}}$ the tension of the string, and $d_H\sim 1/H$ the Hubble length.
The energy density inside the domain wall is of $\order{m_a^2 v_{\rm PQ}^2}$. 
The typical thickness of the domain wall is given by
$m_a(T)^{-1}$~\cite{Vilenkin:1994pv,Linde:1994hy}\ , where 
\begin{align}
	\label{eq:Axion mass}
	m_a(T)^2\simeq \frac{\sqrt{2}\lambda}{m!} \frac{\mpl^3}{v_{\mathrm{PQ}}} \qty[\frac{2(n-3)(n!)^2}{n(n-1)^2\lambda_s^2}]^{\frac{m}{2(n-1)}}  \qty(\frac{H}{\mpl})^{\frac{m}{n-1}}
\end{align}
is induced by the $P$-$S$ mixing.
The typical radius of the cosmic string is given by $v_{\rm PQ}^{-1}$ for $\lambda_p = \order{1}$~\cite{Vilenkin:2000jqa}, with 
the energy density inside the cosmic string of $v_{\rm PQ}^4$.
From these quantities, we obtain,
		\begin{align}
			\label{eq:Tensions of string and wall}
			&\sigma_{\mathrm{w}}\simeq 8m_a(T)v_{\mathrm{PQ}}^2\ , \notag \\
			&T_{\mathrm{s}}\simeq 2\mpi v_{\mathrm{PQ}}^2\ln(v_{\mathrm{PQ}}/H(T))\ .
		\end{align}
By plugging Eq.\,\eqref{eq:Tensions of string and wall} into Eq.\,\eqref{eq:Domain wall collapsing condition}, 
the condition Eq.\,\eqref{eq:Domain wall collapsing condition} is reduced to
\begin{align}
	\label{eq:m_a v.s. H log}
	m_a(T)>\frac{\mpi}{4}H(T)\ln\qty(\frac{v_{\mathrm{PQ}}}{H(T)})
	\ .
\end{align}
 The string-wall network immediately shrinks and collapses when this condition is satisfied 
 for some temperature after $P$ obtains the VEV.
 
It should be noted that the phase component of $S$ also participates in  
the scalar potential induced by the $P$-$S$ mixing. 
Unlike the axion, however, it does not move during the above process.
Thus, we can neglect the motion of the phase component of $S$ in the above argument.
To see this behavior, let us decompose $S$ field by
\begin{align}
	S=\frac{\chi}{\sqrt{2}}\exp\qty(i\frac{b}{\chi})\ ,
\end{align}
where $b$ is the phase component of $S$.
The field value of the radial component $\chi$ slowly decreases 
according to the scaling solution (see Eqs.\eqref{eq:VEV of S in oscillation era} and \eqref{eq:Behavior of S}).
The scalar potential of $a$ and $b$ induced by the $P$-$S$ mixing term is given by,%
\footnote{Here, we redefine the origin of $b$ so that the minimum of the potential is at $a = b = 0$.}
\begin{align}
	\label{eq:Potential of a and b}
	V(a,b)=2\frac{\lambda}{m!\mpl^{m-3}}\qty(\frac{\chi}{\sqrt{2}})^m\frac{v_{\mathrm{PQ}}}{\sqrt{2}}
	\left[1-\cos\qty(\frac{a}{v_{\mathrm{PQ}}}+m\frac{b}{\chi})\right]\ .
\end{align}
This scalar potential implies that the axion oscillates much faster than $b$ 
since $\chi \gg v_{\rm PQ}$ at $T\sim v_{\rm PQ}$.
Thus, $a$ oscillates around $-mbv_{\rm PQ}/\chi$ (mod $2\pi$), 
while $b$ does not feel the force from the 
potential since it is averaged out by the oscillation of $a$.%
\footnote{More precisely, $b$ settles to 
$b_i - m a_i v_{\rm PQ}/ \chi + \order{b_i m^2v_{\rm PQ}^2/\chi^2}$.}

%------------------------
\begin{figure}[t]
		\begin{minipage}{0.49\hsize}
			\centering
			\includegraphics[width=0.88
		\hsize]{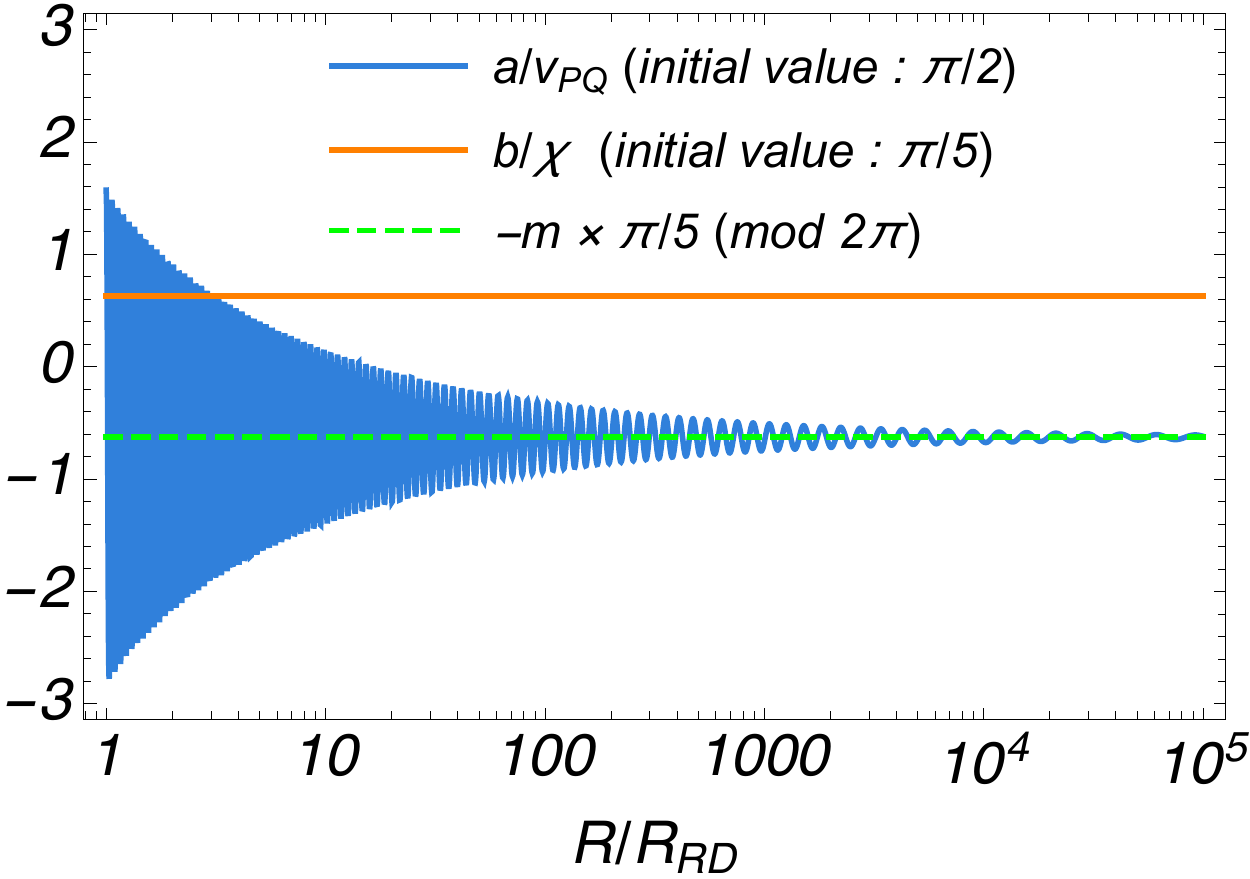}
		\end{minipage}
		\begin{minipage}{0.49\hsize}
			\centering
			\includegraphics[width=0.88
			\hsize]{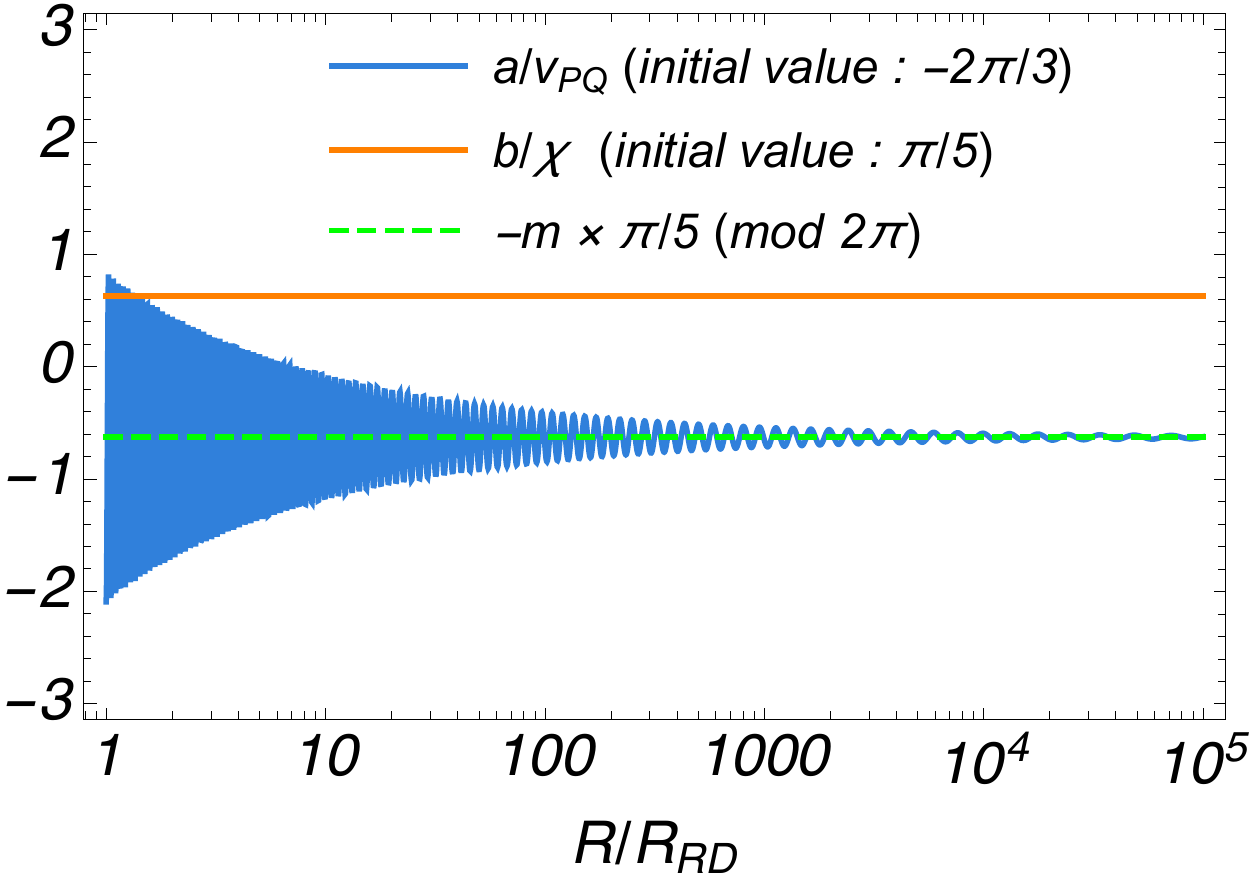}
		\end{minipage}\\
\caption{The behaviors of $a$ and $b$ 
during the RD era at a benchmark point,
$n = 6$, $m=11$, $\lambda_p = c_s = c_p = 1$,
$\lambda_s=10^-4$, $\lambda = 10^{-6}$
and $v_{\rm PQ}=10^{12}$\,GeV.
The initial conditions of the phase components 
are $a/v_{\rm PQ} = \pi/2$ (left)
and $a/v_{\rm PQ} = -3 \pi/2$
while $b/\chi = \pi/5$
in the both panels.
In both cases, $a/v_{\rm PQ}$ settles around 
$-m\pi/5\, ({\mbox{mod}\,2\pi})=- \pi/5$,
while $b/\chi$ is a constant in time.
}
\label{fig:phasemotion}
\end{figure}

In Fig.~\ref{fig:phasemotion},
we show the behaviors of $a$ and $b$ 
during the RD era at a benchmark point,
$n = 6$, $m=11$, $\lambda_p = c_s = c_p = 1$,
$\lambda_s=10^{-4}$, $\lambda = 10^{-6}$
and $v_{\rm PQ}=10^{12}$\,GeV
(see subsection \ref{sec: constraints}).
We start analysis just 
after $P$ obtains the VEV
at $T\sim 10^{12}$\,GeV.
$\chi$ is assumed to follow the scaling solution in Eq.\,\eqref{eq:Behavior of S}.
As the initial conditions of the phase components,
we take $a/v_{\rm PQ} = \pi/2$ (left)
and $a/v_{\rm PQ}= -2\pi/3$ (right),
while $b/\chi = \pi/5$, respectively.
The initial velocities are taken to be zero.
The figure shows that $a/v_{\rm PQ}$
settles around $-mv_{\rm PQ}b/\chi $ (mod $2\pi$).
The figure also shows that $b/\chi$ is an almost constant in time as expected. 
The oscillation period of $a/v_{\rm PQ}$ becomes longer at the later time,
as $\chi$ decreases according to the scaling solution.
In this way, the random field value of $a$ in each Hubble patch settles in 
the vicinity of the uniform field value, $-mbv_{\rm PQ}/\chi$ (mod $2\pi$). 
 
\subsection{The Axion Dynamics After the Onset of $S$ Oscillation }
\label{sec: Axion ramdomize}
As we have mentioned earlier, $S$ starts to oscillate 
around its origin when Hubble parameter becomes smaller than $m_s$.
In this subsection, we consider the dynamics of the axion after the onset of the $S$ oscillation.
\begin{figure}
\centering
	\includegraphics[width=0.7\linewidth]{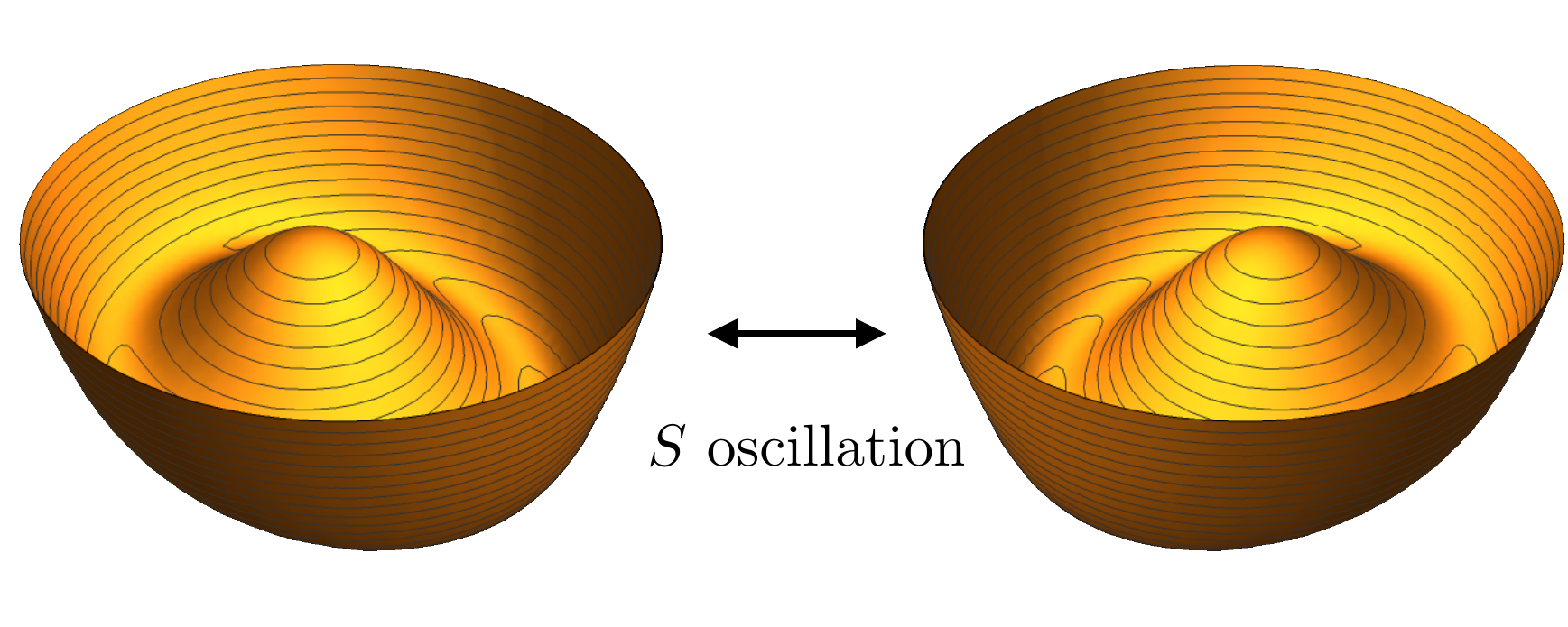}
	\caption{A schematic picture of the behavior of the
	scalar potential of the PQ field.
	As the spectator PQ field oscillates, the sign
	of the axion potential flips.}
	\label{fig:potential_oscillation}
\end{figure}

Because the axion potential is induced by the mixing term, 
the sign of the axion potential also flips
when $S$ oscillates (see Fig.\,\ref{fig:potential_oscillation}).
If the axion mass exceeds the Hubble parameter at that time, the axion falls from the top of the flipped potential. 
If this happens, the axion field value in each Hubble patch 
is again randomized and can no longer be uniform in our universe.
Such a behaviour brings back the domain wall problem.
	
To avoid this situation, we require that the Hubble friction on the axion is effective when $S$ starts to oscillate.
This condition can be expressed in the form
\begin{align}
	\label{eq:start of S oscillation}
	m_a(T)<3H=m_s\ ,
\end{align}
where we estimate the onset of the $S$ oscillation by $3H=m_s$.%
\footnote{Below the temperature of $m_a(T) \simeq 3H$, the field value of $a$ 
is frozen to a different field value in each Hubble patch. 
The difference of the frozen filed value, however, does not cause 
the domain wall problem as long as all the difference is smaller than $2\pi f_a$.}

As we have seen in Eq.\,\eqref{eq:m_a v.s. H log}, the axion mass should exceed the Hubble parameter when $P$
obtains the VEV.
On the other hand, the axion mass needs to be smaller than the Hubble friction when $3H=m_s$.
These two requirements lead to the condition that the axion mass in Eq.\,\eqref{eq:Axion mass},  $m_a\propto H^{\frac{m}{2(n-1)}}$, must decrease 
faster than Hubble parameter.
Thus, the above condition is satisfied for
\begin{align}
	\label{eq:Constraint on n and m}
	m\geq 2n-1\ .
\end{align}

To this point, we have ignored the back reaction to $S$
from $P$. 
As we have seen, $\langle S_I \rangle \gg \langle P_I \rangle$ during inflation.
Thus, the back reaction from $P$ through the mixing term
is negligible during inflation.
As the field value of $P$ decreases much faster than that of $S$, the back reaction is also negligible in the 
inflaton oscillation era and the RD era until $P$ obtains 
the VEV.
	
After $P$ obtains the VEV, 
$\langle P \rangle \simeq v_{\mathrm{PQ}}$, 
the back reaction
could modify the behavior of $S$ in Eq.\,\eqref{eq:Behavior of S} 
if
\footnote{Here, we use the lower limit on $m$ in Eq.\,\eqref{eq:Constraint on n and m}
since the back reaction is weaker for a larger $m$.}
\begin{align}
    \label{eq:backreaction}
		\frac{\lambda_s^2}{(n!)^2\mpl^{2n-4}} \abs{S} ^{2n} &\sim \frac{\lambda}{(2n-1)!\mpl^{2n-4}}  \abs{S}^{2n-1}v_{\mathrm{PQ}}\notag \,\\
		\therefore S &\sim \frac{(n!)^2\lambda}{(2n-1)!\lambda_s^2}v_{\mathrm{PQ}}.
\end{align}
However, such a situation can be avoided if
the mass of $S$ is larger so that $S$ starts oscillation 
before Eq.\,\eqref{eq:backreaction} is satisfied.
Thus, so long as
\begin{align}
	\label{eq:Constraints from back reaction}
		\lambda v_{\mathrm{PQ}}\mpl\qty(\frac{(n!)^2\lambda}{m!\lambda_s^2}\frac{v_{\mathrm{PQ}}}{\mpl})^{2n-3}<m_s^2\ ,
\end{align}
we can safely neglect the back reaction of $P$ to the dynamics of $S$.

%=======================%
%	Cosmological Constraints   %
%=======================%
%=========================
\subsection{Isocurvature Perturbations of the Axion}
\label{sec:Isocurvature Perturbation of Axion}
Because $P$ obtains the VEV during 
the RD era, it seems that the axion does not suffer from the 
isocurvature problem.
However, the phase component $b$ of $S$ has a flat potential 
when $S$ takes a large field value, and it obtains 
quantum fluctuation during inflation.
In the presence of the mixing term, the fluctuation of 
the phase component of $S$ is imprinted in the axion, 
which leads to the isocurvature perturbations of the axion dark matter.

During inflation, the fluctuation of $b$ is given by~\cite{Mukhanov:1981xt,Hawking:1982cz,Starobinsky:1982ee,Guth:1982ec,Bardeen:1983qw},
\begin{align}
	\delta b_I\simeq\frac{H_I}{2\pi} \ ,
\end{align}
where $\langle \chi_I \rangle = \sqrt{2}\langle S_I \rangle$.%
\footnote{Since $P$ also takes a large field value, $a$ also fluctuates during inflation.
However, the axion eventually 
settles around $-mv_{\rm PQ}/\chi b$ regardless of its initial value as we have seen in subsection~\ref{subsec:After PQ Breaking}. 
Thus, the fluctuation of $a$ does not affect the following arguments.} 
After inflation, $S$ follows the scaling solution which is along 
the straight line passing through $S=0$ in the complex plane of $S$.
Thus, the fluctuation of $\delta b$ decreases as
	\begin{align}
		\delta b\simeq\frac{\chi}{\langle{\chi_I}\rangle} \times \delta b_I
		\simeq 
		\frac{H_I}{2\mpi}
		\frac{\chi}{\langle \chi_I \rangle}
		\ ,
	\end{align}

Once $P$ obtains the VEV, the $P$-$S$ mixing term leads to
the potential of $a$ and $b$ in Eq.\,\eqref{eq:Potential of a and b}.
Then, as we discussed in subsection~\ref{subsec:After PQ Breaking},
$a/v_{\rm PQ}$ settles around $-m\times b/\chi $
(mod $2\pi$) while $b/\chi$ does not move.
As a result, the fluctuation of $b$ is imprinted in $a$ as%
\footnote{Only the fluctuation modes longer than the Hubble length at the QCD temperature 
are relevant for the isocurvature perturbations of the axion dark matter, which 
are superhorizon mode when $a$ settles around $-m v_{\rm PQ}/\chi b$.}
\begin{align}
\label{eq:dela}
    \frac{\delta a}{f_a} \simeq m\frac{v_{\rm PQ}}{\chi} \frac{\delta b}{f_a} 
    \simeq \frac{m N_{\rm DW} H_I}{2\pi \langle\chi_I\rangle}\ .
\end{align}

Below the QCD scale, the axion starts coherent oscillation 
and the axion fluctuation results in the uncorrelated isocurvature perturbations.
The power spectrum of the isocurvature perturbations is given by
\begin{align}
\label{eq:isocurvature}
    {\cal P}_I \simeq 8\left( \frac{\delta a}{f_a}\right)^2
    \left(\frac{f_a}{10^{12}\,\rm GeV}\right)^{1.19}
    \left(\frac{\Omega_ah^2}{0.1}\right)\ ,
\end{align}
where we assume that the observed dark matter density is dominated by the axion.
From the CMB observations, the uncorrelated isocurvature perturbations of cold dark matter are constrained by \cite{Akrami:2018odb},
\begin{align}
\label{eq:isoconst}
    \beta_{\rm iso} = \frac{{\cal P}_I}{{\cal P}_\zeta+{\cal P}_I} \le 0.038\ ,
\end{align}
where ${\cal P}_\zeta \simeq 2\times 10^{-9}$ denotes the power spectrum of the curvature perturbations.
By combining Eqs.\,\eqref{eq:VEV of S in inflation era}, \eqref{eq:dela} and \eqref{eq:isocurvature},
we find the upper limit on $H_I$ is given by,
\begin{align}
    H_I \lesssim \mpl \times \left(
    8\times 10^{-10} \frac{1}{m^2 N_{\rm DW}^2}
     \left(\frac{10^{12}\,\rm GeV}{f_a}\right)^{1.19}
    \left(\frac{0.1}{\Omega_ah^2}\right)
    \right)^{\frac{n-1}{2(n-2)}}\qty(\sqrt{\frac{c_s}{n}}\frac{n!}{\lambda_s})^{\frac{1}{n-2}}
\end{align}

\subsection{Viable Parameter Region}
\label{sec: constraints}
	\begin{figure}[t]
		\begin{minipage}{0.49\hsize}
			\centering
			\includegraphics[width=0.8\hsize]{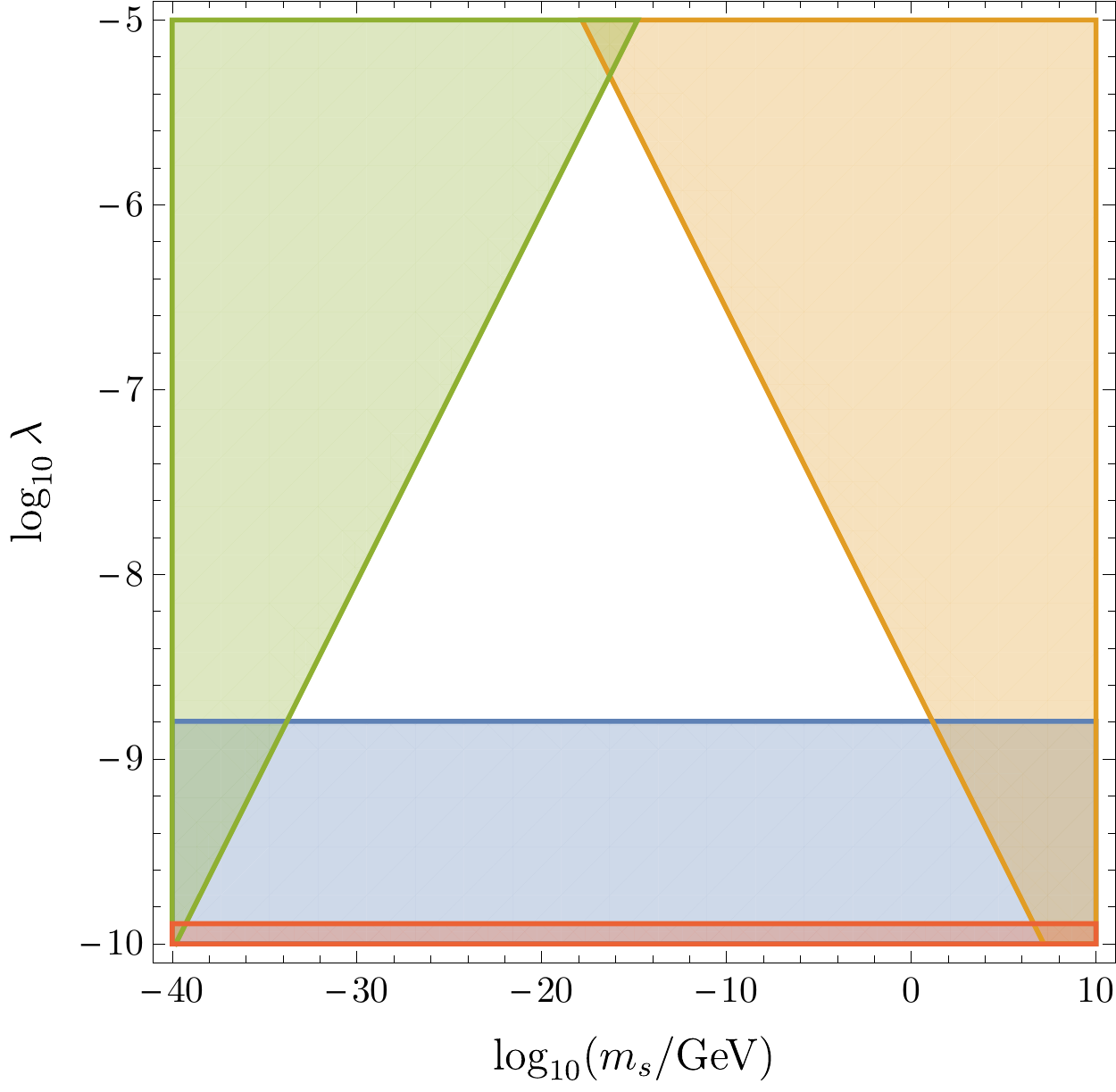}
		\end{minipage}
		\begin{minipage}{0.49\hsize}
			\centering
			\includegraphics[width=0.8\hsize]{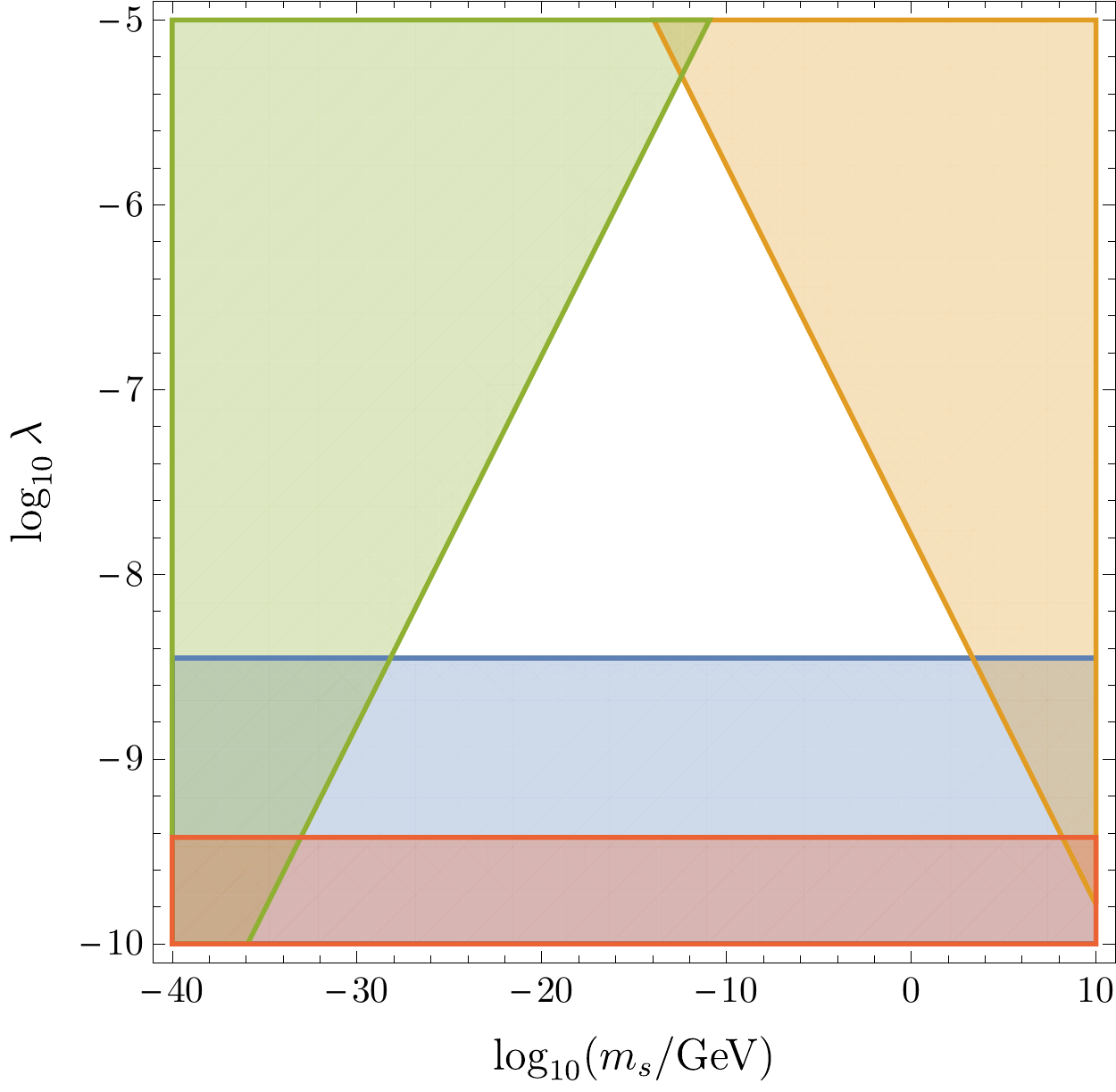}
		\end{minipage}\\
\caption{Various constraints on the model parameters. The blue, orange and green shaded areas are excluded by the conditions in Eqs.\,\eqref{eq:m_a v.s. H log}, \eqref{eq:start of S oscillation}, \eqref{eq:Constraints from back reaction}.
The red one is excluded by a conservative condition $m_a(T\sim v_{\rm PQ}) \gtrsim 3H(T\sim v_{\rm PQ})$.
As a benchmark, we set $n=6$, $m=11$, and  $\lambda_p=c_p=c_s=1$ and $\lambda_s = 10^{-4}$.
{\it Left}) $N_{\mathrm{DW}}=1$,
{\it Right}) $N_{\mathrm{DW}}=6$.}
\label{fig:totalconstn6m11}
\end{figure}

Let us summarize the constraints on the model parameters. In Fig.~\ref{fig:totalconstn6m11}, we show the constraints on the $(m_s, \lambda)$ plane,
for  $N_{\rm DM}=1$ (the minimal KSVZ axion model)
and for $N_{\rm DM}= 6$ (DFSZ axion model~\cite{Zhitnitsky:1980tq,Dine:1981rt}).%
\footnote{Here, we assume that the two Higgs doublets in the DFSZ model couple to
the PQ field via $P^2H_1 H_2$.}
We take $n=6$, $m=11$, $\lambda_p = 1$,
$c_s = 1$ and $c_p = 1$ as a benchmark point.
The blue shaded region is excluded where the condition 
in Eq.\,\eqref{eq:m_a v.s. H log} is not satisfied 
after $P$ obtains the VEV, and hence, the randomness of the 
axion direction is not resolved.
The orange shaded region is excluded 
where the condition in Eq.\,\eqref{eq:start of S oscillation}
is not satisfied at $3H \simeq m_s$,
and hence, the axion is randomized as $S$ oscillates.
The green shaded region is excluded 
where the condition in Eq.\,\eqref{eq:Constraints from back reaction} is not satisfied,
where the back reaction from $P$ to $S$ becomes sizable.
The red shaded region shows a more conservative 
constraint, $m_a(T\sim v_{\rm PQ}) \gtrsim 3H(T\sim v_{\rm PQ})$, 
which is weaker than that in Eq.\,\eqref{eq:m_a v.s. H log}.
This weaker constraint is good enough if no cosmic strings are formed 
when $P$ obtains the VEV at $T\sim v_{\rm PQ}$.
The figure shows that all the conditions are satisfied for a wide range of $m_s$.
For the minimal KSVZ model, i.e. $N_{\rm DW} =1$,
$m_s \lesssim 1$\,GeV, and for 
the DFSZ model, $m_s\lesssim 10^2$\,GeV.

For the benchmark scenario, $n=6$, $m=11$,
the constraint on the isocurvature perturbations leads to
\begin{align}
    H_I \lesssim 10^{12}\,{\rm GeV}\times 
    \frac{1}{N_{\rm DW}^{5/4}} \left(\frac{\sqrt{c_s}}{\lambda_s}\right)^{1/4} \ .
\end{align}
Thus, we find that the present mechanism 
allows $H_I \sim 10^{13}$\,GeV 
for $\lambda_s \sim 10^{-6}$.%
\footnote{For $\lambda_s \sim 10^{-6}$,
the expectation value of $S$ during inflation 
slightly exceeds $\mpl$, which requires a small $c_s$
to avoid too much potential energy of $S$ during inflation.}

As a result, we find that the present mechanism
makes the axion model with $N_{\rm DM} \neq 1$ 
compatible with the inflation model 
in which a large Hubble parameter is rather large, i.e., $H_{I} \gg 10^{7\mbox{--}8}$\,GeV.
Therefore, this mechanism increases the freedom of choice
of the combinations of the axion models
and inflation models.

%===========%
%	SUSY ver.   %
%===========%
%=========================
\section{Supersymmetric Realization}
\label{sec:SUSY}
As we have seen in the previous section, the model 
requires the higher dimensional interaction terms
with specific exponents.
Such scalar potentials are not easily justified 
in non-supersymmetric theories.
In this section, we briefly discuss a supersymmetric realization of the mechanism to make the scenario 
more viable.
A detailed analysis of the supersymmetric extension will be given elsewhere.

For example, a model with $n = 6$ and $m=11$ can be
easily realized by assuming a superpotential,
	\begin{align}
	\label{eq:superpotential}
		W=X(P\bar{P}-v_{\mathrm{PQ}}^2)+\frac{1}{\mpl^4}YS^6+Z\qty(S^2P+\frac{1}{\mpl^6}S^9)+\frac{1}{\mpl^6}X\bar{P}S^7
		\ .
	\end{align}
\color{black}Here, $P$ and $S$ are the chiral superfields 
corresponding to the (spectator) PQ fields, 
while $X$, $Y$ and $Z$ are chiral superfields whose $F$-components lead to the scalar potential in Eq.\,\eqref{eq: Potential of P,S}.%
\footnote{In this realization, the $P$-$S$ mixing term in the scalar potential
is given by $|S|^{4}S^7P^* + \hc$
Accordingly, the cosine potential of the phase components in Eq.\,\eqref{eq:Potential of a and b} 
is modified to $\cos(a/v_{\rm PQ}+ 7b/\chi)$.}
We omit the coupling constants for brevity.
The PQ charge assignment is given in Table~\ref{table: SUSY charge assignment}.
This model justifies why the potential term with 
a lower dimension than $|S|^{2n}$ are absent.
The mass term of the scalar component of $S$ is 
generated by the supersymmetry breaking effects.
It should be noted that the unwanted superpotential 
terms such as $YP^{i}S^j$ with $i>1$ $(i + j = 6)$
are suppressed by PQ and R-symmetry.

\begin{table}[t]
	\begin{center}
		\caption{Charge assignments of the supersymmetric model.
		In addition to the PQ symmetry,
		we show the $R$-symmetry.
		}
		\label{table: SUSY charge assignment}
		\begin{tabular}{|c|c|c|c|c|c|c|c|c|c|}
			\hline 
			& $X$ & $Y$ & $Z$ & $S$ & $P$ & $\bar{P}$ & $Q_L\bar{Q}_R$ &
			$\psi_s\bar{\psi}_s$  \\ 
			\hline 
			$U(1)_{\rm PQ}$ & $0$ & $-6$ & $-9$ &$1$ & $7$ & $-7$ &$-7$ &$-1$   \\ 
			\hline
			$R$ & $2$ & $2$ & $2$ & $0$ & $0$ & $0$ & $2$ & $2$  \\
			\hline 
		\end{tabular} 
	\end{center}
\end{table}

The last term of Eq.\,\eqref{eq:superpotential} induces
    \begin{align}
        V= \frac{1}{\mpl^6}(P\bar{P}-v_{\mathrm{PQ}}^2)\bar{P}^*S^7+\hc
    \end{align}
When $T\gtrsim v_{\mathrm{PQ}}$, $P$ and $\bar{P}$ are settled to the origin due to thermal mass terms, thus this term does not affect the dynamics of $S$ and $P$.
In addition, after $\langle P\bar{P}\rangle$ settles to $v_{\mathrm{PQ}}^2$, this term vanishes.
Therefore, this term does not have an influence on the dynamics all through the epoch of interest.

We briefly comment on the effects of higher dimensional terms in the K\"ahler potential.
For example, a higher dimensional operator, 
    \begin{align}
    \label{eq:Higher Kahler1}
        K = \frac{1}{\mpl^2}\abs{X}^2\abs{S}^2\ , 
    \end{align}
induces a scalar potential,
    \begin{align}
    \label{eq:potential from Kahler}
        V = \frac{|P\bar P- v_{\rm PQ}^2|^2}{1+|S|^2/\mpl^2 }\ .
    \end{align}
This term leads to an additional effective mass of $S$ for a given $\langle P\bar{P}\rangle$.
During inflation, this term is smaller than the Hubble induced mass term (see Eqs.\,\eqref{eq:P vev inflation} and \eqref{eq:P vev inflation num}). After inflation,  $\langle P\bar{P}\rangle$ immediately vanishes, and the mass term in Eq.\,\eqref{eq:potential from Kahler} leads to a mass of $\order{v_{\rm PQ}^2/\mpl}$, which is smaller than the 
Hubble constant until $T \sim v_{\rm PQ}$.
After $\langle P \bar{P}\rangle$ settles to $v_{\rm PQ}^2$,
the induced mass vanishes.%
\footnote{In the presence of the supersymmetry breaking effects, the VEV of $P\bar{P}$ is slightly shifted from $v_{PQ}^2$, and hence,
the effective mass in Eq.\,\eqref{eq:potential from Kahler} does not vanish completely.
However, we can set this mass term small enough not to affect the dynamics.
}
Therefore, we find that the induced mass does not affect 
the dynamics of the scalar fields.

In addition to the higher dimensional term in Eq.\,\eqref{eq:Higher Kahler1}, there are terms
    \begin{align}
     K = \frac{1}{\mpl^2} (\abs{Y}^2\abs{S}^2 + \abs{Z}^2\abs{S}^2 + \abs{S}^4) + \cdots\ .
    \end{align}
These terms just induce additional terms with higher dimension than those in Eq.\,\eqref{eq: Potential of P,S}, and hence, they do not affect the scaling behavior of $S$.%
\footnote{We assume that $X,Y,Z$ obtain the positive Hubble mass terms, hence they settle at the origin. }
In the case of $|S|^4$, we redefine $S$ so that it has a canonical kinetic term.
With the redefinition, the effects appear as the additional terms with higher dimension than those in Eq.\,\eqref{eq: Potential of P,S}. The same is true for the higher dimensional K\"ahler term of $|S|^{2k}$ $(k > 1)$.

Supersymmetric extension is also advantageous 
to explain the interactions between the PQ fields and 
the inflaton in Eq.\,\eqref{eq: Potential of P,S and inflaton}.  
In fact, a K\"ahler potential,
\begin{align}
\label{eq:Kahler of I S P}
K = \frac{c_p}{3\mpl^2} \abs{I}^2  \abs{P}^2   
- \frac{c_s}{3\mpl^2} \abs{I}^2  \abs{S}^2   \ ,
\end{align}
explains the interactions between the PQ fields and the inflaton.
Here, $I$ denotes the chiral superfield of the inflaton $\phi$
whose $F$-component provides the inflaton potential, 
i.e., $ V(\phi) = |F_I|^2 $.

In addition to Eq.\,\eqref{eq:Kahler of I S P}, we can write higher order K\"ahler terms such as
    \begin{align}
        K = \abs{I}^2\qty(\frac{\abs{S}^4}{\mpl^4}+\frac{\abs{S}^6}{\mpl^6}+\cdots).
    \end{align}
In the case of $\langle S_I \rangle\sim \mpl$, these terms slightly modify the expectation value of $S$ during inflation.
However, after inflation, $\langle S \rangle$ starts to decrease, then the higher order contributions become negligible, therefore they do not affect the dynamics.

\section{Conclusions}
The domain wall problem and the isocurvature 
problem restrict possible combinations
of axion models and inflation models.
In this paper, we considered a new mechanism 
which solves those problems by introducing 
the spectator PQ field which obtains 
a large field value before the PQ field obtains 
the VEV.
The mechanism makes the axion model with a non-trivial domain wall number
compatible with the inflation model 
with a large Hubble parameter, 
$H_{I} \gg 10^{7\mbox{--}8}$\,GeV.
The mechanism is also free from the isocurvature problem.
It should be emphasized that this mechanism can be 
added to any conventional axion models.
Thus, this mechanism increases the freedom of choice
of combinations of axion models and inflation models.

We also find that the present mechanism can be consistent with a large Hubble parameter during inflation, of $H_I\sim 10^{13}$\,GeV.
Thus, the scenario can be tested by combining future axion search experiments and the searches for the primordial 
$B$-mode polarization in the CMB. 

The model also predicts the existence of the spectator PQ field.
As we have discussed in \ref{subsec:Radiation Dominated Era},
the coherent oscillation of the spectator PQ field 
can play a role of the dark matter when it is very light.%
\footnote{In this case, we do not need to introduce 
the light fermions in Eq.\,\eqref{eq:Decay channel of S}.}
As the initial amplitude of the coherent oscillation is 
dynamically determined, this model realizes 
a very light scalar dark matter without fine-tuning of the initial condition in an alternative way to the axion-like
ultra-light dark matter in \cite{Hu:2000ke,Hui:2016ltb}.
Such a very light dark matter can be tested via 
astronomical ephemeris~\cite{Fukuda:2018omk}.

%==================%
%	Acknowledgements   %
%==================%
%=========================
\begin{acknowledgements}
The authors thank M.~Kawasaki for an important comment on the isocurvature problem.
This work is supported in part by JSPS Grant-in-Aid for Scientific Research No. 16H02176 (T.T.Y), No. 17H02878 (M.I., and T.T.Y.), No. 15H05889, No. 16H03991, No. 18H05542 (M.I.) and by World Premier International Research Center Initiative (WPI Initiative), MEXT, Japan (M.I., and T.T.Y.). 
T.T.Y. is supported in part by the China Grant for Talent Scientific Start-Up Project. T.T.Y. thanks to Hamamatsu Photonics.
The work of M.S. is supported in part by a Research Fellowship for Young Scientists from the Japan Society for the Promotion of Science (JSPS). 
\end{acknowledgements}
%=========================

%=============%
%	Appendices    %
%=============%
%=========================

\appendix
%%%%%%%%%%%%%%%%%%%%%%%%%
\section{Energy Density of the Spectator PQ Field}\label{sec:energy density of S}
In this appendix, we discuss the 
energy density of the spectator PQ field, $S$.
When $S$ follows the scaling solution in 
Eqs.\,\eqref{eq:VEV of S in inflation era}, \eqref{eq:VEV of S in oscillation era} and \eqref{eq:Behavior of S},
the potential energy density of $S$ is of ${\cal O}(H^2 |S|^2)$.
Thus, it is sub-dominant compared 
with the dominant energy density of ${\cal O}(H^2 \mpl^2)$
as long as $\langle S\rangle \ll \mpl$.

Once $S$ starts coherent oscillation around its origin, $S$ behaves as a massive matter 
with an energy density 
\begin{align}
    \rho_S = m_s^2 |S|^2\ .
\end{align}
The radiation density at that time is $m_s^2 \mpl^2/3 $ where we have used $H\simeq m_s/3$.
Thus, again, the energy density of $S$ is  sub-dominant since $S\ll \mpl$ at the onset of the coherent oscillation.

As we considered in \ref{sec:model setup}, $S$ immediately decays into the massless fermions
which behave as radiation.
Thus, the energy density of $S$ does not causes any cosmological problems.
As the energy density of $S$ is sub-dominant,
the energy density of the massless fermions 
are also sub-dominant.
Furthermore, the relative entropy of the 
massless fermions is diluted when 
all the entropy in the thermal bath 
goes into the particles in the standard cosmology (i.e. the photons and the neutrinos).
Thus, the contributions of the massless fermions to the dark radiation is also negligible.

%%%%%%%%%%%%%%%%%%%%%%%%%%%%%%%%%%%%%%%%%%%%%%%%
%\bibliographystyle{utphys}
\bibliography{papers}

\end{document}